\documentclass[a4paper,amsmath,amssymb,superscriptaddress,aps,prb,reprint,showpacs]{revtex4-1}
\usepackage{xcolor,graphicx}
\usepackage{subfigure}
\usepackage{MnSymbol}
\usepackage{braket}
\usepackage{tikz}
\usepackage{pgfplots}
\usepackage[utf8]{inputenc}

\newcommand{\ind}{_\mathrm}
\renewcommand{\i}{\mathrm i}
\renewcommand{\exp}[1]{\mathrm{e}^{#1}}
\newcommand{\cdag}{c^{\dagger}}
\newcommand{\cnod}{c^{\phantom{\dagger}}}

\renewcommand{\vec}[1]{\mathbf{#1}}
\renewcommand{\onlinecite}[1]{\cite{#1}}

\begin{document}

\title{Excitonic Magnetism at the intersection of Spin-orbit coupling and crystal-field splitting}

\author{Teresa Feldmaier}
\author{Pascal Strobel}
\affiliation{%
Institute for Functional Matter and Quantum Technologies,
University of Stuttgart,
70550 Stuttgart,
Germany}
\author{Michael Schmid}
\affiliation{%
Institute for Functional Matter and Quantum Technologies,
University of Stuttgart,
70550 Stuttgart,
Germany}
\affiliation{Center for Integrated Quantum Science and Technology, University of Stuttgart,
Pfaffenwaldring 57, 
70550 Stuttgart, Germany}
\author{Philipp Hansmann}
\affiliation{
Max-Planck-Institute for Chemical Physics of Solids, N\"othnitzer Straße 40,
01187 Dresden, Germany}
\affiliation{Department of Physics, University of Erlangen-Nuremberg, 91058 Erlangen, Germany}
\author{Maria Daghofer}
\affiliation{%
Institute for Functional Matter and Quantum Technologies,
University of Stuttgart,
70550 Stuttgart,
Germany}
\affiliation{Center for Integrated Quantum Science and Technology, University of Stuttgart,
Pfaffenwaldring 57, 
70550 Stuttgart, Germany}

\date{\today}

\begin{abstract}
Excitonic magnetism involving superpositions of singlet and triplet states
is expected to arise for two holes in strongly correlated and spin-orbit
coupled $t_{2g}$ orbitals. However, uncontested material examples for its realization are
rare. We apply the Variational Cluster Approach to the square lattice
to investigate excitonic antiferromagnetism and the impact of a
crystal field. We give a phase diagram depending on spin-orbit
coupling and crystal field and find excitonic magnetism to survive in the presence
of substantial crystal-field--induced orbital order. We address the specific example of 
Ca$_2$RuO$_4$ using \textit{ab initio} modeling and conclude it to realize
such excitonic magnetism despite its pronounced orbital
polarization. We also reproduce magnetic excitations and show the maximum at momentum
$(0,0)$ to be related to the excitonic character of the magnetic order. 
\end{abstract}

\maketitle

\section{Introduction} \label{sec:intro}  
Strong correlations in transition-metal oxides have for decades been a focus
of intense scientific activity. While the prominent $3d$ metals, e.g. 
iron, nickel and copper, are usually understood to have rather weak spin-orbit
coupling (SOC), the interplay of strong SOC and correlations has
attracted attention more recently.
One important driver of this interest is the search for topologically nontrivial phases like
topological Mott insulators~\cite{Pesin:2010ju} or spin liquids. In particular, honeycomb-lattice
compounds with a single hole in strongly correlated $t_{2g}$ orbitals of $4d$ (e.g. Ru) or $5d$ (e.g. Ir)
elements are candidates for 'Kitaev' spin liquids~\cite{Kitaev:2006ik} driven by
distinct magnetic anisotropies~\cite{PhysRevLett.102.017205,Chaloupka:2010gi}, which has led to extensive research efforts~\cite{0953-8984-29-49-493002,2019arXiv190308081T}. 

The impact of SOC is particularly transparent for a $t_{2g}$ electron,
where it couples effective orbital angular momentum $L=1$ and spin $S=\tfrac{1}{2}$ to total angular momentum
$J=\tfrac{3}{2}$ or $J=\tfrac{1}{2}$~\cite{giniyat_orb_order_and_fluct}. 
If sufficiently large, this leads to a band splitting and the reduced
width of the resulting subbands effectively enhances correlations. The single-layer  
square-lattice compound Sr$_2$IrO$_4$ has been established as a prime example
for such spin-orbit driven Mott physics~\cite{rev_square_iri}. With five $t_{2g}$ electrons, its
$J=\tfrac{3}{2}$ levels are completely filled and the $J=\tfrac{1}{2}$ level
is half filled. The resulting effective single-band description
focusing on the $J=\tfrac{1}{2}$ states is even robust enough to survive in doped regimes,
where Fermi-arcs~\cite{Kim:2014hn} and hints of
superconductivity~\cite{dwave_Sr2IrO4} have been reported, as well as
photo-doping~\cite{PhysRevX.9.021020}. 
 
\begin{figure}
\includegraphics[width=\columnwidth]{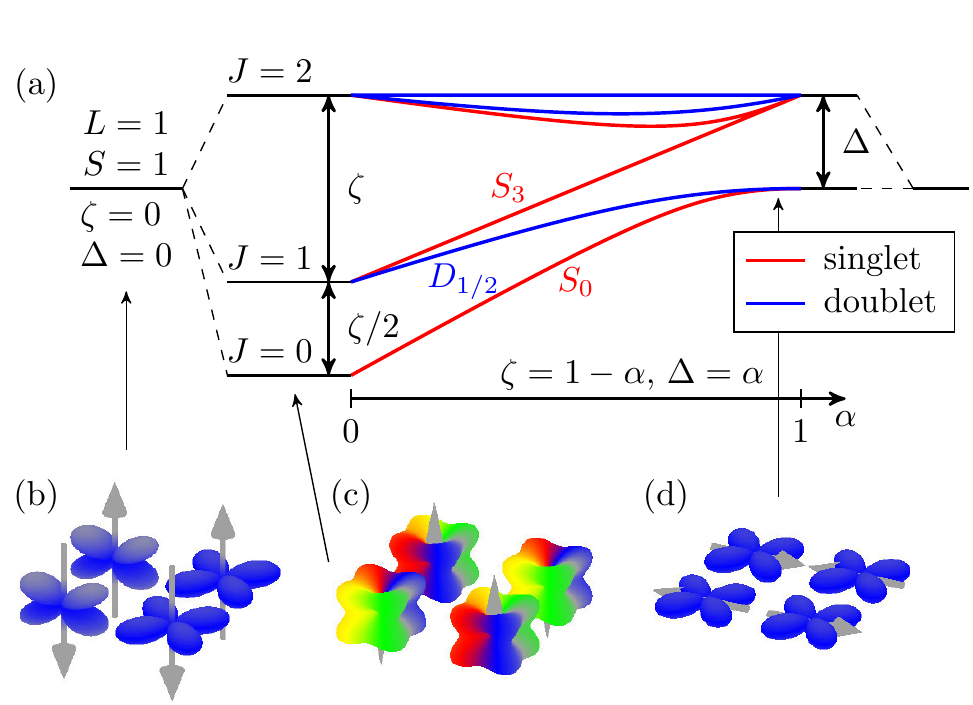}
\caption{(a) Level structure starting from nine-fold degeneracy (far left and
  right) and interpolating between spin-orbit coupled (left, $\alpha=0$) and
  crystal field (CF) split (right, $\alpha=1$) regimes. In the limit
  $\Delta=\alpha=0$ (left), singlet $S_3$ and doublet $D_{1/2}$ are
  degenerate and give the $J=1$ triplet; in the limit
  $\zeta=1-\alpha=0$ (right), ground-state singlet  $S_0$ and
  $D_{1/2}$ correspond to $S=1$ and doubly filled $xy$ orbital. (b) is
  the stripy
  state arising out of nine-fold degeneracy via Goodenough-Kanamori rules. (c)
  Excitonic magnetism for dominant SOC, one state of the $J=1$ triplet 
  condenses to yield a magnetic moment and mixes with the $J=0$
  singlet. (d) Orbitally ordered state with doubly occupied $xy$
  orbital, where holes in $xz$ and $yz$ form  $S=1$. In (b) and (d), the
  doubly occupied orbital is drawn. \label{fig:delta_lambda}
}
\end{figure}

While a description in terms of $J=\tfrac{1}{2}$ is thus well established for
a single hole in the $t_{2g}$ levels of Ru or Ir and forms the basis for the proposed
spin-liquid scenario, the
situation for \emph{two holes} is less clear. Double-perovskite
iridates~\cite{PhysRevB.93.035129,PhysRevLett.120.237204} likely realize a
nonmagnetic singlet groundstate and might already be close to the $j$-$j$--coupling regime
with a doubly occupied $J=\tfrac{1}{2}$
band~\cite{RIXS_Ca2RuO4_Gretarson19}. Theoretical predictions for substantial but not extreme spin-orbit
coupling, where $L$-$S$ coupling is more appropriate, argue that it combines effective angular momentum $L=1$ 
with total spin $S=1$ of the two holes into an ionic ground state with
$J=0$~\cite{Khaliullin:2013du,PhysRevB.91.054412}, see the left-hand
side of Fig.~\ref{fig:delta_lambda}(a). Superexchange then determines
dynamics of  $J=1$ excitations as well as their potential condensation
into excitonic magnetic order. This unconventional type of
magnetism has been predicted to host a bosonic Kitaev-Heisenberg model and topologically
nontrivial excitations~\cite{PhysRevLett.122.177201,PhysRevB.100.224413}, or to mediate triplet
superconductivity~\cite{PhysRevLett.116.017203}.

Ca$_2$RuO$_4$ is a promising candidate for such a scenario, where superexchange
mixes  $J=0$ and $J=1$ states so that the superposition acquires a magnetic moment that can order, see the schematic
illustration in Fig.~\ref{fig:delta_lambda}(c). Features like $XY$-symmetry
breaking (rather than Ising or Heisenberg) and the corresponding maximum of
excitation energies at momentum $(0,0)$ 
can be explained by such an excitonic model~\cite{Higgs_Ru,PhysRevLett.119.067201}. However, a structural phase
transition is well established~\cite{PhysRevLett.95.136401,PhysRevLett.87.077202}, which leads to a
crystal field (CF) lowering the  energy of the $xy$ orbital as shown on the right-hand side of Fig.~\ref{fig:delta_lambda}(a). 
Based on {\it ab initio}
calculations~\cite{PhysRevB.95.075145,hund_Ca2RuO4},  the two holes are
distributed mostly in $xz$ and $yz$ orbitals, suggesting more
conventional $S=1$ spin order, see the sketch of
Fig.~\ref{fig:delta_lambda}(d). SOC 
would then be a correction rather than a main driver~\cite{PhysRevB.95.075145,PhysRevLett.115.247201}. 

We apply here the variational cluster approach (VCA) to a spin-orbit
coupled and correlated $t_{2g}$ three-orbital model at nominal filling $n=4$ in order to address the
competition between SOC and CF as illustrated in
Fig.~\ref{fig:delta_lambda}. We find that without CF splitting, SOC $\zeta$ suppresses orbital order, so that the complex spin and orbital pattern of
Fig.~\ref{fig:delta_lambda}(b) is replaced by the excitonic antiferromagnetic (AFM) state 
Fig.~\ref{fig:delta_lambda}(c). In the presence of CF splitting,
its impact will be seen to be more gradual, but it nevertheless induces a transition from
spin order, as in Fig.~\ref{fig:delta_lambda}(d), to excitonic order. In order to connect
to Ca$_2$RuO$_4$, we use a realistic model based on density-functional theory
and subsequent projection to Wannier states~\cite{PhysRevLett.123.137204}. The approach is validated
by the magnetic excitation
spectrum of a spin-orbital model based on the same parameters, which
reproduces neutron-scattering data, and leads to the conclusion that
the compound falls into the excitonic regime despite its strong
orbital polarization. 

In Sec.~\ref{sec:model}, we introduce the $t_{2g}$ model with SOC  and also
give the Kugel-Khomskii-like superexchange model valid in the limit of strong
onsite interactions in Sec.~\ref{sec:kk}. We also shortly mention the VCA. We
then give results in Sec.~\ref{sec:results}, first for the orbitally
degenerate model in Sec.~\ref{sec:delta_0} and then in the presence of a
crystal field in Sec.~\ref{sec:delta}, where we also discuss
Ca$_2$RuO$_4$. Finally, we and provide magnetic
excitation spectra in Sec.~\ref{sec:Skw} and conclude in Sec.~\ref{sec:conclusions}.

\section{Model and Methods} \label{sec:model}
We discuss four electrons (two holes) in a three-orbital system modeling $t_{2g}$ orbitals on a square
lattice. The kinetic energy
\begin{align}
H\ind{kin}= \sum_{\vec{k},\sigma} \sum_{\alpha,\beta}
\epsilon_{\alpha,\beta}(\vec{k})\cdag_{\vec{k},\alpha,\sigma}\cnod_{\vec{k},\beta,\sigma}\;,
\end{align}
where $\cnod_{\vec{k},\alpha,\sigma}$ ($\cdag_{\vec{k},\alpha,\sigma}$) annihilates
(creates) an electron with spin $\sigma=\uparrow,\downarrow$ in orbital
$\alpha=xy,xz,yz$ and momentum $\vec{k}$, has the following contributions:
\begin{align}\label{Hkinxy}
  \epsilon_{xy,xy}(\vec{k}) &= -2t_{xy}^{\textrm{NN}} (\cos k_x + \cos k_y) \nonumber\\
  &\quad - 4t_{xy}^{\textrm{NNN}}\cos k_x\cos k_y-\Delta\;,
\end{align}
where tetragonal CF splitting $\Delta > 0$ originates from the strained octahedral coordination of the
low-temperature phase of Ca$_2$RuO$_4$,
\begin{align}\label{Hkinxz}
\epsilon_{xz,xz}(\vec{k}) &= -2t_{xz}^{\textrm{NN}} \cos k_x,\quad
\epsilon_{yz,yz}(\vec{k}) = -2t_{yz}^{\textrm{NN}} \cos k_y,
\end{align}
and
\begin{align}\label{Hkinto}
\epsilon_{xz,xy}(\vec{k}) = -2t^{o} \cos k_x\;,
\end{align}
where orbital-mixing $t^o$ arises from the orthorhombic distortion. 
The SOC operator for $t_{2g}$ orbitals can be written as \cite{PhysRevB.73.094428,PhysRevB.98.205128} 
\begin{equation}
  H\ind{SOC}=\zeta \sum_i \vec{l}_{i} \cdot \vec{s}_{i}
=
\frac{i\zeta}{2}\sum_i\sum_{\stackrel{\alpha,\beta,\gamma}{\sigma,\sigma'}} 
\varepsilon^{\phantom{\alpha}}_{\alpha\beta\gamma} \tau^\alpha_{\sigma\sigma'}
\cdag_{i,\beta,\sigma}\cnod_{i,\gamma,\sigma'}\;,
  \label{HSOC}
\end{equation}
where $\varepsilon_{\alpha\beta\gamma}$ is the totally antisymmetric
Levi-Civita tensor and $\tau^\alpha$ with $\alpha=x,y,z$ are Pauli matrices. 

The effective onsite two particle interaction is included in the $SU(2)$ symmetric Kanamori aproximation~\cite{PhysRevB.28.327} 
\begin{align} \label{Hint}
H\ind{int} &= U \sum_{i, \alpha} n_{i \alpha \uparrow} n_{i \alpha \downarrow} 
  +\frac{U^\prime}{2} \sum_{i, \sigma} \sum_{\alpha \neq \beta} n_{i \alpha
    \sigma} n_{i \beta \bar{\sigma}}\\ \nonumber
    & +\frac 1 2 (U^\prime - J_H) \sum_{i,\sigma} \sum_{\alpha \neq \beta}
  n_{i \alpha \sigma} n_{i \beta \sigma}\\ \nonumber
    & - J_H \sum_{i, \alpha \neq \beta} ( c^\dagger_{i \alpha \uparrow}
  \cnod_{i \alpha \downarrow} c^\dagger_{i, \beta \downarrow} \cnod_{i \beta
    \uparrow}- c^\dagger_{i \alpha \uparrow} c^\dagger_{i \alpha \downarrow} \cnod_{i \beta \downarrow} \cnod_{i \beta \uparrow})
\end{align}
with intraorbital Hubbard $U$, interorbital $U^\prime$, and Hund's coupling $J\ind H$
which are related by $U^\prime = U-2J\ind H$.

\subsection{Variational Cluster Approximation}\label{sec:VCA}

In order to address long-range order and symmetry breaking, we use the
VCA~\cite{PhysRevLett.91.206402}, which has already been established to
explain angle-resolved photo emission for strongly spin-orbit coupled
Sr$_2$IrO$_4$~\cite{PhysRevLett.105.216410}. VCA is complementary to 
static mean-field methods, which cannot describe the large-$U$ limit of
$t_{2g}$ orbitals with SOC correctly~\cite{kaushal2020bcsbec}, and dynamical mean-field theory
with Monte-Carlo impurity solvers which are restricted to regimes above the N\'eel
temperature due to the minus-sign
problem~\cite{PhysRevLett.123.137204}. The VCA is further not restriced to the
one-dimensional limit addressed with the density-matrix renormalization
group~\cite{PhysRevB.96.155111,kaushal2020bcsbec}.

In our VCA simulations, the  self energy of a $2\times 2$-sites cluster, with three $t_{2g}$ orbitals per
site, is calculated using exact diagonalization (ED) and then inserted into the
one-particle Green's function of the thermodynamic limit to obtain the grand
potential. According to self-energy functional theory~\cite{Pot03}, the thermodynamic grand potential $\Omega[\vec
  \Sigma_{\vec \tau^\prime}]$ can be optimized by varying the
one-particle parameters $\vec{\tau}^\prime$  used to obtain the cluster self energy $\Sigma_{\vec \tau^\prime}$. 
In particular, the self energy is optimized w.r.t. fictitious
symmetry-breaking fields, e.g. a staggered magnetic field inducing AFM order. It should be noted that these fields only act on the reference system, they are not
included in the thermodynamic-limit Green's functions. If, however, a  symmetry-broken self energy optimizes the grand
potential of the fully symmetric Hamiltonian, one can infer spontaneous symmetry breaking. 


\begin{table}
  \caption{Ficticious Weiss fields $\Lambda$ and ordering momenta
    $\vec{Q}$ as in Eq.~(\ref{eq:weiss}), as well as combinations
    thereof used to describe various phases.\label{tab:patterns}}
  \begin{ruledtabular}
    \begin{tabular}{|l|cccc|}
      & $\Lambda_1$ & $\vec{Q}_1$ & $\Lambda_2$ & $\vec{Q}_2$  \\ \hline
    stripy $S_z$ & $S_z$ & $(\pi,0)$ & &  \\
    stripy $S_z$, stripy orb.& $S_z$ & $(\pi,0)$ & $n_{xy}-n_{xz}$ &
               $(0,\pi)$ \\
    & \multicolumn{2}{c}{$\Lambda_3=n_{yz}$}  & \multicolumn{2}{c|}{$\vec{Q}_3=(0,0)$}\\
    stripy $S_x$& $S_x$ & $(\pi,0)$ & &  \\
    stripy $S_x$, stripy orb.& $S_x$ & $(\pi,0)$ & $n_{xy}-n_{xz}$ &
    $(0,\pi)$ \\
    & \multicolumn{2}{c}{$\Lambda_3=n_{yz}$}  & \multicolumn{2}{c|}{$\vec{Q}_3=(0,0)$}\\
    $G$-AFM $S_z$& $S_z$ & $(\pi,\pi)$ &&\\
    $G$-AFM $S_x$& $S_x$ & $(\pi,\pi)$ &&\\
    $G$-AFM $S_z$, FO $xy$& $S_z$ & $(\pi,\pi)$ & $n_{xy}$ & $(0,0)$ \\
    $G$-AFM $S_z$, TO& $S_z$ & $(\pi,\pi)$ & \multicolumn{2}{c|}{orb. see
    Ref.~\cite{PhysRevLett.88.017201}} \\
    $G$-AFM $2S_z + L_z$& $2S_z + L_z$ & $(\pi,\pi)$&&\\
    $G$-AFM $2S_x + L_x$& $2S_x + L_x$ & $(\pi,\pi)$&&\\
    $G$-AFM $M_z$& $2S_z - L_z$ & $(\pi,\pi)$&&\\
    $G$-AFM $M_x$& $2S_x - L_x$ & $(\pi,\pi)$&&\\
    $G$-AFM $J_z$& $S_z + L_z$ & $(\pi,\pi)$&&\\
    $G$-AFM $J_x$& $S_x + L_x$ & $(\pi,\pi)$&&\\
    $G$-AFM $S_x+S_y$& $S_x + S_y$ & $(\pi,\pi)$&&\\
    \end{tabular}
  \end{ruledtabular}
\end{table}

We sampled a variety of potential ordering parameters, which can generally be
expressed as one-particle terms of the form
\begin{equation}\label{eq:weiss}
H_h = h^\prime \sum_{i} \Lambda_i \exp{\i \vec Q \vec R_i}\;, 
\end{equation} 
where $\Lambda_i$ can be any one-particle operator acting on site $i$ at
position $\vec R_i$, e.g. spin $S^\alpha_{i}$ along direction $\alpha$, total
angular momentum $S^\alpha_{i}+L^\alpha_{i}$, magnetization
$2S^\alpha_{i}-L^\alpha_{i}$ or any other linear combination of $S$ and
$L$, as well as 
orbital density $n_{i,\beta}$ in orbital $\beta$. Ordering  vectors $\vec Q$ accessible to our cluster
are $(0,0)$, $(\pi,\pi)$, $(\pi,0)$ and $(0,\pi)$. We also included
superpositions of two or three order parameters, e.g., orbital and
magnetic order, with identical or different (e.g. in the stripy phase)
ordering vectors, see Tab.~\ref{tab:patterns}. The patterns tested were motivated on one hand by the states
giving a large contribution to the ED ground state, and on the other hand by
phases reported in the literature for similar models~\cite{PhysRevLett.88.017201,PhysRevB.74.195124}.

In addition to these fictitious fields, we had two more variational
parameters, namely the physical chemical potential $\mu$ and the fictitious
$\mu'$. The chemical potential $\mu$ ensures a particle
density of 2 holes (4 electrons) per site and is implemented by the 
requirement
\begin{align}
  \frac{\partial F}{\partial \mu} = \frac{\partial \Omega + \mu N_{\textrm{target}}}{\partial \mu} =0\;
\end{align}
with $N_{\textrm{target}}=2$ (in hole notation)~\cite{PhysRevB.82.174441}. Together with the chemical-potential term
\begin{align}\label{eq:musumn}
-\mu N  := -\mu\sum_{i,\alpha,\sigma} n_{i,\alpha,\sigma}\;,
\end{align}
contained in the Hamiltonian, where the sum runs over over site $i$, orbital $\alpha$ and spin
$\sigma$, stationarity translates to $N_{\textrm{target}} = \langle N\rangle$.
However, the actual overall density $\langle N\rangle_{\mu^*}$ obtained  for
the saddle-point chemical potential $\mu^*$ does not necessarily
fulfill $\langle N \rangle_{\mu^*} = N_{\textrm{target}}$, and we indeed found it to differ slightly. To ensure
thermodynamic consistency~\cite{AichhornEPL72_117}, we thus additionally optimized the
fictitious chemical potential $\mu'$. Like the symmetry-breaking
fields $h'$, it acts only on the reference system. The robust
Mott gap present in all considered parameter regimes ensures a very weak
dependence of $\mu$ and $\mu'$ on the various $h'$, so that we fixed
$\mu$ and $\mu'$ to the values found for $h'=0$ and then kept them constant
during optimization of the $h'$.

One-particle observables like orbital- or spin-resolved densities can be
obtained from the thermodynamic-limit one-particle Green's function that is calculated in the
VCA. Two-particle quantities are not as straightforward, we approximate them
here as the expectation value w.r.t the  ED ground
state  $|\phi_0\rangle$  calculated 
with optimized order parameters. In particular, we are interested in the
weight found in eigenstates $|J,J^z\rangle$ with total onsite angular momentum $J=0,1,2$,
i.e. in
\begin{align}\label{eq:wght_j}
\langle J \rangle := \frac{1}{N_{\textrm{sites}}}\sum_{i,J^z}\langle \phi_0| J_i,J^z_i \rangle \langle J_i,J^z_{i}
|\phi_0\rangle\;,
\end{align}
where we average over the $N_{\textrm{sites}}$ sites $i$ of the
$2\times 2$ cluster. Analogously, one can define weight in other
onsite two-hole states, e.g., the states $S_0$, $D_{1/2}$, $S_3, \dots$ of the level structure shown in
Fig.~\ref{fig:delta_lambda}(a). 

\subsection{Perturbation theory and Kugel-Khomskii--type model}\label{sec:kk}

While the VCA provides one-particle 
spectral functions, two-particle quantities like the dynamic magnetic
structure factor are not readily available. We thus resort to ED
on a small cluster. In order to go beyond the $2\times 2$ cluster, we
restrict the Hilbert space to the nine low-energy  states of Fig.~\ref{fig:delta_lambda}. 
In this low-energy manifold, the two holes
on each site reside in two different orbitals, i.e. we conveniently label the
orbital wave function by the remaining doubly occupied orbital. This gives us the effective
orbital angular momentum $L=1$~\cite{Khaliullin:2013du,PhysRevLett.89.167201}. Additionally, Hund's-rule coupling enforces a
total spin $S=1$, leading to a total of nine low-energy states. The VCA
results show that these nine states indeed capture  nearly all the
weight of the cluster ground state, which justifies projecting out charge
excitations as well as states with onsite spin $S=0$. The restricted
Hilbert space then allows us to reach eight rather than four sites in exact diagonalization. 

The effective Hamiltonian acting on this low-energy manifold is obtained by treating hopping $t$ in second-order
perturbation theory, with the large energy scale being onsite interactions
(\ref{Hint}), associated with charge excitations and local singlets.
We give here the various terms of
the effective Kugel-Khomskii--type Hamiltonian, restricting
ourselves to orbital-conserving hoppings for simplicity, i.e., we neglect
interorbital $t^o$ which would significantly complicate the superexchange model. The first possible
process is a spin-spin coupling 
\begin{align}
H_{\vec{S}\cdot\vec{S}} = J_{i,j}\left( \vec{S}_i\vec{S}_j - 1\right)\otimes |T_i;T_j\rangle\langle T_i;T_j|
\end{align}
that leaves orbital occupations $T_i$ and $T_j$ 
unchanged. Its coupling strength depends on whether the same orbital $\gamma$
is doubly occupied on both sites, or whether it is orbital $\alpha$ on one
site and orbital $\beta$ on  the other, i.e.,
\begin{align}
J_{i,j} = \begin{cases}
(t_\alpha^2 + t_\beta^2)\frac{U+J_H}{U(U+2J_H)}\ &\textrm{for}\ T_i=T_j=\gamma\\
\frac{t_\gamma^2(U+J_H)}{U(U+2J_H)}-\frac{J_H(t_\alpha^2 + t_\beta^2)}{U(U-3J_H)} \ &\parbox{0.18\textwidth}{for $T_i=\alpha \neq
T_j=\beta$ and $\alpha,\beta\neq \gamma$}
\end{cases}
\end{align}
and accordingly
\begin{align}\label{eq:HSS_s}
  H_{\vec{S}\cdot\vec{S}}& = \left( \vec{S}_i\vec{S}_j - 1\right)
  \biggl[\sum_\gamma \bigl(\sum_{\alpha\neq \gamma}t_\alpha^2\bigr)
    \frac{U+J_H}{U(U+2J_H)} |\gamma;\gamma\rangle\langle
    \gamma;\gamma|\\ \label{eq:HSS_d}
    &+ \sum_{\substack{\alpha \neq \beta\\ \gamma\neq\alpha,\beta}} \Bigl(\frac{t_\gamma^2(U+J_H)}{U(U+2J_H)}-\frac{J_H(t_\alpha^2 + t_\beta^2)}{U(U-3J_H)}\Bigr)
    |\alpha;\beta\rangle\langle \alpha;\beta|\biggr]\;.
\end{align}
The first of these would be the only superexchange term surviving in a
strongly orbitally polarized case for  crystal field $\Delta\gg
\zeta$.

Additional terms involve the orbital degree of freedom. Again
discussing first the case of identical orbital character on both
sites, finite Hund's-rule coupling $J_H > 0$ (or rather the associated
pair-hopping term) allows the doubly occupied orbital
to change  on both sites simultaneously, as long as
not all spins are parallel, yielding a 'pair-flip' term
\begin{align}\label{eq:Hp}
  H_{p}& = \left( \vec{S}_i\vec{S}_j - 1\right) \Bigl[\sum_{\alpha\neq \beta} 
      \frac{- t_\alpha t_\beta J_H}{U(U+2J_H)}
      |\alpha;\alpha\rangle\langle \beta;\beta|\Bigr]\;.
\end{align}

The counterpart for different orbital occupation on both sites gives
a  diagonal part, in addition to the diagonal
terms already included in $S_i^zS_j^z$ above, that is independent of the
$S^z$ components on the two sites and reads
\begin{align}\label{eq:difforb_diag}
\langle S_i^z, T_i=\beta; S_j^z,T_j=\gamma | H |S_i^z, T_i=\beta;
S_j^z,T_j=\gamma\rangle = -\frac{t_\beta^2+t_\gamma^2}{U-3J_H}\;.
\end{align}
This diagonal term can be combined with orbital and spin-orbital--flip terms into 
\begin{align}\label{eq:HTT_s}
  H_{\vec{TT}}&= -\sum_{\beta\neq \gamma}\frac{t_\beta^2+t_\gamma^2}{U-3J_H}|\beta;\gamma\rangle\langle\beta;\gamma|\\
  &\quad+\sum_{\beta\neq \gamma}\frac{t_\beta
  t_\gamma}{U(U-3J_H)}[2J_H+(U-J_H)\bigl(
  \vec{S}_i\vec{S}_j+1\bigr)] \nonumber\\\label{eq:HTT_d}
&\qquad\qquad\otimes|\beta;\gamma\rangle\langle\gamma;\beta|\;. 
\end{align}

The full superexchange Hamiltonian for a pair $(i,j)$ of sites combines (spin-)orbital exchange (\ref{eq:HTT_s}) and
(\ref{eq:HTT_d}) with Heisenberg spin interaction Eqs.~(\ref{eq:HSS_s}) and
(\ref{eq:HSS_d}) as well as (\ref{eq:Hp}) into 
\begin{align}\label{eq:HKKij}
H_{i,j} =   H_{\vec{S}\cdot\vec{S}} + H_p + H_{\vec{TT}}\;.
\end{align}
For each nearest-neighbor (NN) pair $\langle i,j \rangle$, two of the three hoppings
$t_{\alpha/\beta/\gamma}$ are non-zero (see main text), for next-nearest
neighbors $\llangle i,j \rrangle$, only the $xy$ orbital hops.

These intersite coupling terms due to hopping are complemented with the onsite
SOC and CF. Crystal field $\Delta$ from Eq.~(\ref{Hkinxy}) favors the $xy$ orbital to be doubly
occupied, which is chosen as the $L^z=0$ state, and SOC Eq.~(\ref{HSOC}) favors the $J=0$
state, so that the onsite Hamiltonian becomes
\begin{align}\label{eq:ham_onsite}
  H_{\textrm{ion}} = \frac{\zeta}{2} \vec{S}\vec{L} + \Delta (L^{z})^2\;
\end{align}
after projection onto the low-energy $S=1$ manifold. Eigenstates of this
onsite Hamiltonian define the level structure shown in Fig.~\ref{fig:delta_lambda}.

The translation between the notation $|\beta\rangle$ giving the doubly
occupied orbital $\beta=yz,xz,xy$, as
used in the super-exchange model above, and the  orbital
angular momentum operators is achieved by~\cite{Khaliullin:2013du,PhysRevLett.89.167201}
\begin{align}\label{eq:L}
  L^x &= i\left(|xy\rangle\langle xz|-|xz\rangle\langle xy|\right)\\
  L^y &= i\left(|yz\rangle\langle xy|-|xy\rangle\langle yz|\right)\\
  L^z &= i\left(|xz\rangle\langle yz|-|yz\rangle\langle xz|\right)\;.
\end{align}
The operator $(L^{z})^2 = |xz\rangle\langle xz|+|yz\rangle\langle yz|$ thus
indeed favors doubly occupied $|xy\rangle$ states. 

We use Lanczos exact diagonalization to obtain spectra 
\begin{align} \label{eq:Mkw}
  M^{\alpha}(\vec{k},\omega) = -\frac{1}{\pi} \Im \langle \phi_0 |M^{\alpha}(-\vec{k})
  \frac{1}{\omega - H +\i0^+}M^{\alpha}(\vec{k})|\phi_0\rangle\;,
\end{align}
where $M^{\alpha}(\vec{k}) = \tfrac{1}{N} \sum_i
\textrm{e}^{i\vec{k}\vec{r}_i} M^\alpha_{i}$ for component $\alpha =a,b,c$ of
$\vec{M}=2\vec{S}-\vec{L}$, and effective low-energy $H$ composed of onsite and intersite terms
(\ref{eq:ham_onsite}) and (\ref{eq:HKKij}). We are able to treat an eight-site cluster of $\sqrt{8}\times\sqrt{8}$
geometry.

\section{Results}\label{sec:results}

\begin{figure}
\includegraphics[width=0.9\columnwidth]{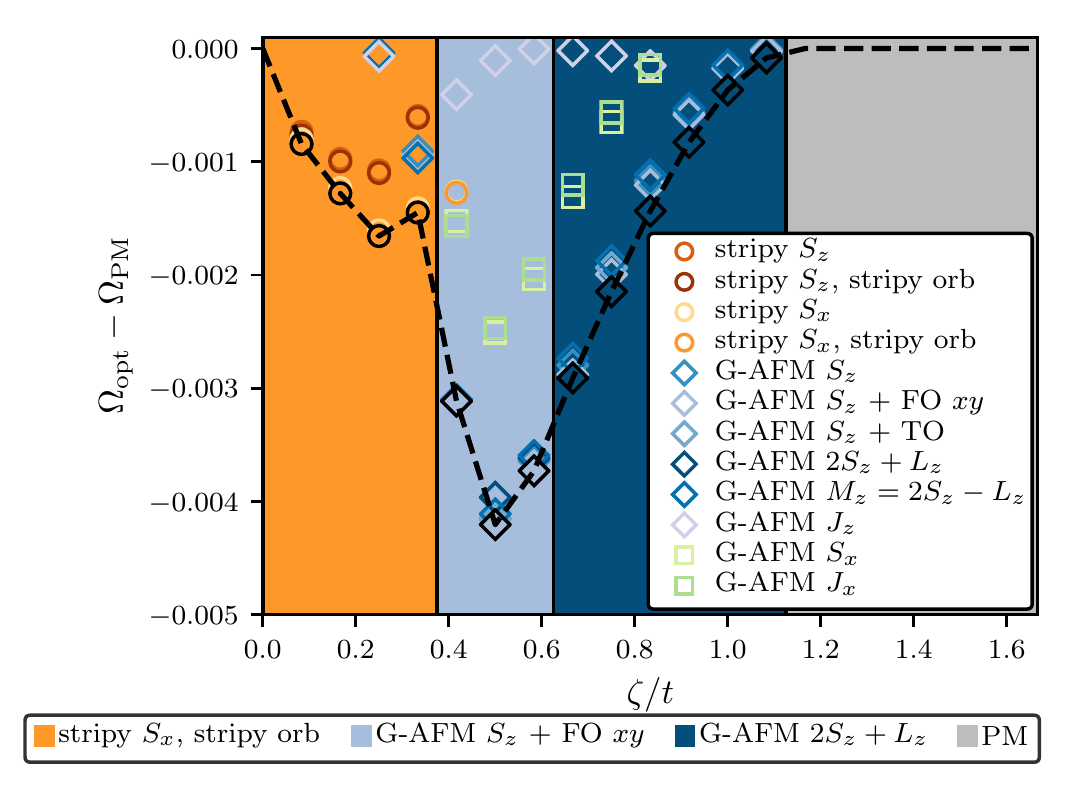}
\caption{Grand canonical potential per site obtained with VCA for various
  order parameters in units of $t$. $\Omega_{\textrm{PM}}$ is the
  value obtained without symmetry breaking and $\Omega_{\textrm{opt}}$
  is the lowest grand potential found by allowing for various ordered
  states, see Tab.~\ref{tab:patterns}: '$G$-AFM' refers to checkerboard pattern, 
  i.e. $\vec{Q}=(\pi,\pi)$ in Eq.~(\ref{eq:weiss}), 'stripy' to
  $\vec{Q}=(0,\pi)$, and 'FO' to ferrorobital order [$\vec{Q}=(0,0)$]
  favoring the $xy$ orbital, 'TO' is the complex three-orbital--order
  of Ref.~\onlinecite{PhysRevLett.88.017201}. 'Stripy orb.' denotes
  the pattern of Fig.~\ref{fig:delta_lambda}(b), with
  $\vec{Q}_{\textrm{orb}}=(\pi,0)$ orthogonal to
  $\vec{Q}_{\textrm{spin}}=(0,\pi)$ of the concurrent stripy spin
  pattern.   The dashed line is a guide to the
  eye connecting the lowest-$\Omega$ values, whose color is used for
  the background. A gray background denotes the paramagnetic regime,
  where no symmetry breaking lowers the grand potential. Parameters are
  $t_{xy}^{\textrm{NN}} =t_{xz}^{\textrm{NN}} =t_{yz}^{\textrm{NN}} = t$,
  $\Delta=0$, $U=12.5t$ and $J_H=2.5t$.\label{fig:omega} }
\end{figure}

\begin{figure}
\includegraphics[width=\columnwidth]{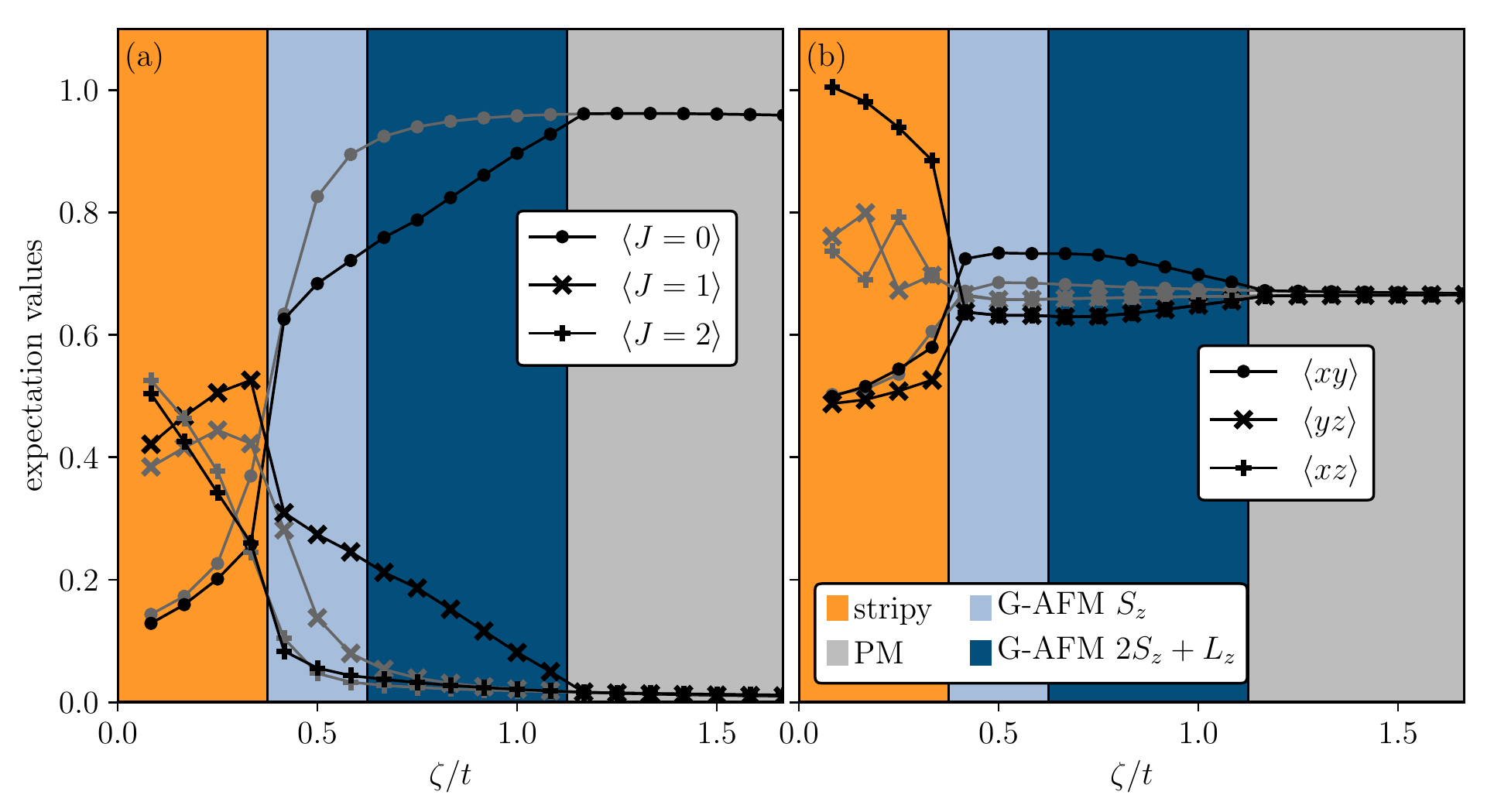}
\caption{Spin-orbital onsite state and orbital occupations for
  parameters as in Fig.~\ref{fig:omega}, i.e. without
  crystal field. Gray symbols give (a) weights Eq.~(\ref{eq:wght_j}) found in states
  with  total onsite angular momentum  $J=0, 1, 2$ and (b) orbital
  occupation numbers $n^h_{yz/xz/xy}$ for holes. Black symbols indicate
  the same observables, but obtained for the ordered state, i.e. with
  optimized symmetry-breaking cluster parameters as inferred from Fig.~\ref{fig:omega}.  
  Background color labels the ordered phase yielding the lowest grand
  potential as in Fig.~\ref{fig:omega}: orange - stripy [see Fig.~\ref{fig:delta_lambda}(b), but with
    in-plane spins], blue - checkerboard $G$-AFM order with out-of-plane moment (light
  blue: $S_z$ plus a small $xy$ polarization; dark blue: $2S_z+L_z$), and gray -
  paramagnet.  
 \label{fig:delta_0}}
\end{figure}

We are first going to discuss a more generic model for excitonic
magnetism with and without CF splitting, where we
focus on essentials. We choose $t = t_{xy}^{\textrm{NN}} =
t_{xz}^{\textrm{NN}} =t_{yz}^{\textrm{NN}}$ as unit of energy and set
$t_{xy}^{\textrm{NNN}} = t^{o} = 0$. Interactions are fixed to 
$U=12.5t$ and $J_H=2.5t$, which is consistent with their order of magnitude in
Ca$_2$RuO$_4$~\cite{hund_Ca2RuO4} and solidly in the Mott insulating
regime. SOC $\zeta$ and CF $\Delta$ will be varied in order to assess their
impact. 

\subsection{Excitonic AFM ordering without CF}\label{sec:delta_0}

Figure~\ref{fig:omega} shows how much the VCA grand potential $\Omega$
is lowered by symmetry breaking w.r.t. the paramagnetic (PM) phase, for various potential order
parameters. Crystal field is here $\Delta=0$ and the ground-state ordering for each $\zeta$ is the one giving the lowest
$\Omega$. For $\zeta >0$, the out-of-plane $z$-component and the
in-plane $x$-component of angular momentum lead to clearly different
values, reflecting the loss of rotational symmetry. Different linear combinations of $S$ and $L$, on the other hand, 
typically only give slightly different grand potentials  --
the crucial aspect of the order parameter is the symmetry broken
rather than the exact form of the order parameter.

For vanishing or small SOC, the lowest grand potential is obtained for a complex pattern
with orbital and spin order having orthogonal ordering vectors $(0,\pi)$ and
$(\pi,0)$, as depicted in Fig.~\ref{fig:delta_lambda}(b). Such a state is consistent with
Goodenough-Kanamori rules, has also been
found by ED~\cite{PhysRevB.74.195124}, and remains stable for $\zeta \lesssim 0.4t$. The only role of SOC is
here a slight preference for $S^x$ over $S^z$. 
At $\zeta\approx 0.4t$, the magnetic ordering vector switches to
$(\pi,\pi)$ in a first-order transition, see Fig.~\ref{fig:omega}, and orbital
densities become more equal as stripy orbital order is lost, see Fig.~\ref{fig:delta_0}(b).

With the transition to checkerboard order, total onsite angular momentum
$J$ becomes more relevant, as can be inferred from the weight
(\ref{eq:wght_j}) in eigenstates $|J=0,1,2\rangle$. The $J=0$ state
quickly gains weight with the onset of checkerboard AFM order, as shown in
Fig.~\ref{fig:delta_0}(a) and similarly seen using dynamical mean-field
theory~\cite{PhysRevB.99.075117}. In fact, it almost entirely describes the
PM state already for $\zeta \approx 0.6t$, see the gray symbols
in Fig.~\ref{fig:delta_0}(a).
Magnetic ordering increases weight in $J=1$ (see black symbols), exactly as expected for
excitonic magnetism, while the $J=2$ are nearly irrelevant in the entire
checkerboard regime. The small orbital
polarization $n_{xy}^h > n_{xz}^h, n_{yz}^h$, see Fig.~\ref{fig:delta_0}(b) is due to
the layered geometry and favors an ordered moment along $z$ rather than in plane. 
The onset of checkerboard order together with the shift to the $J=0$ state suggests that
ordering vector $(\pi,\pi)$ is favored as soon as $\zeta$ sufficiently lifts
orbital degeneracy.

\subsection{Competition of SOC and  CF}\label{sec:delta}

\begin{figure}
\includegraphics[width=0.9\columnwidth]{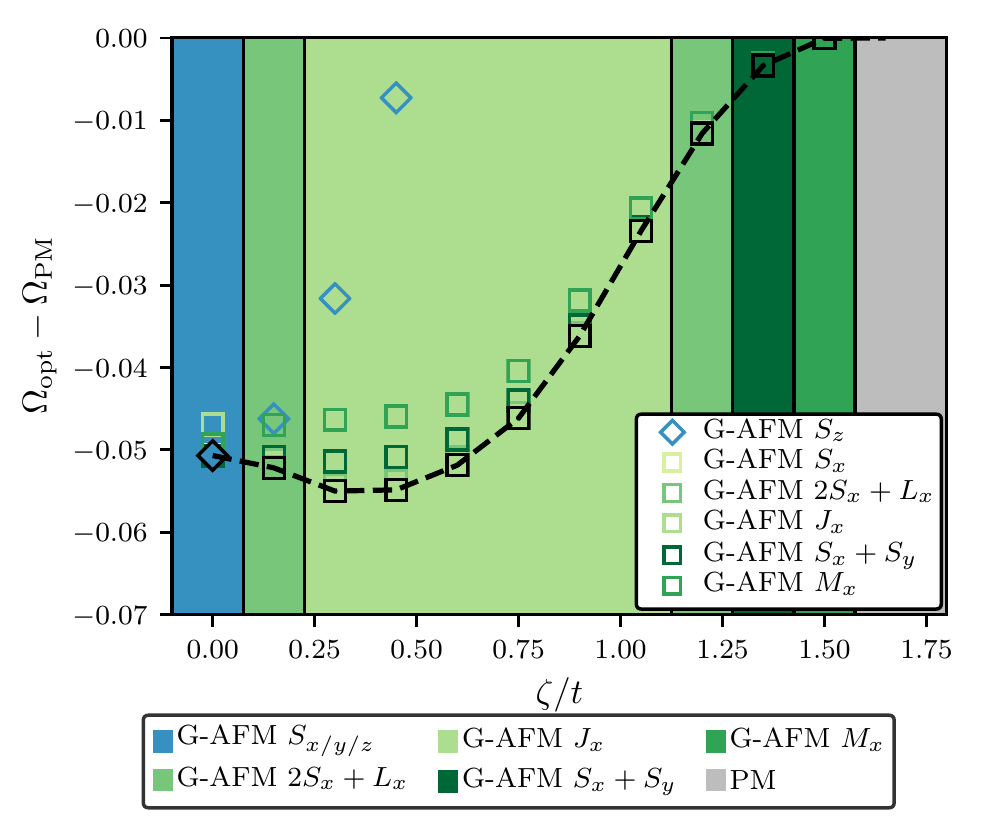}
\caption{Grand canonical potentials obtained with VCA with crystal
  field $\Delta=1.5t$, remaining parameters as in Fig.~\ref{fig:omega}.
  All magnetic ordering vectors are
  $\vec{Q}_{\textrm{magn}}=(\pi,\pi)$ and the substantial CF suppresses any orbital order, so that
  corresponding Weiss fields are all 0.\label{fig:omega_delta} }
\end{figure}

\begin{figure}
\includegraphics[width=\columnwidth]{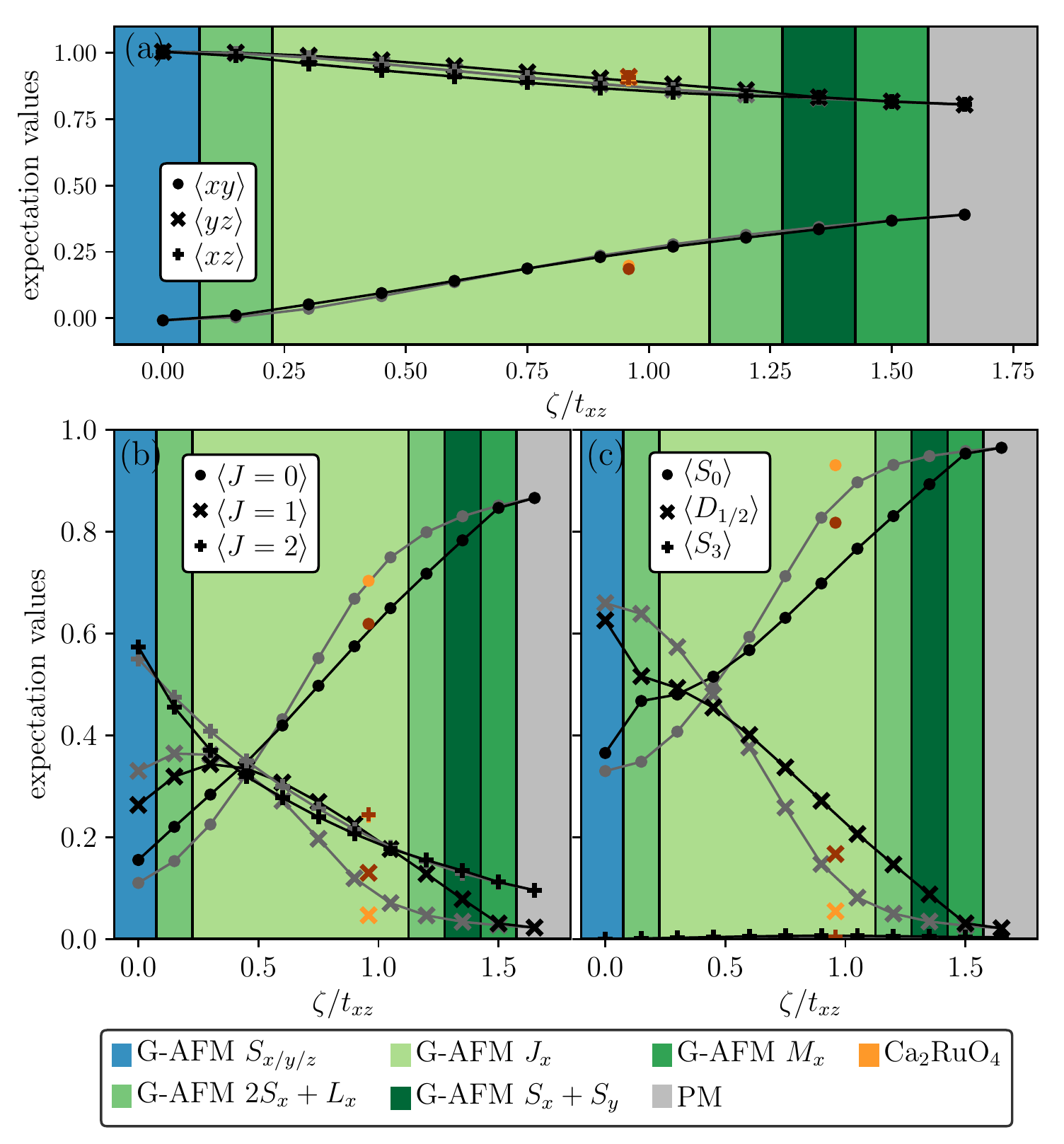}
\caption{Orbital occupation and spin-orbital order in the presence of a CF. (a)  gives orbitally resolved hole occupation numbers
  $n^h_{yz/xz/xy}$, (b) weights (\ref{eq:wght_j}) found in states
  with  total onsite angular momentum  $J=0, 1, 2$, and (c) shows analogous weights in
  eigenstates of (\ref{eq:ham_onsite}), i.e., in the levels $S_0$, $D_{1/2}$,
  and $S_3$ of 
  Fig.~\ref{fig:delta_lambda}(a). Gray/black symbols are results obtained without/with AFM
  order. Background color refers to the optimal order parameter inferred from
  the VCA grand potentials of Fig.~\ref{fig:omega_delta} and  given below
  the figure; it  always
  lies in the $x$-$y$ plane, except for the isotropic
  limit $\zeta=0$. Parameters as in Fig.~\ref{fig:omega} except that 
  $\Delta=1.5t$. Orange/red symbols refer to the parameters used to model 
  Ca$_2$RuO$_4$, see text, and the state without/with $G$-AFM symmetry breaking,
  the latter prefers moments along $b=\tfrac{1}{\sqrt{2}}(x+y)$
  direction. 
\label{fig:delta}}
\end{figure}

Having established the signatures of excitonic magnetism in the case of
degenerate orbitals, we now include a tetragonal CF splitting that favors
doubly occupied $xy$ orbitals and thus breaks orbital degeneracy from the
outset. This is the situation depicted in Fig.~\ref{fig:delta_lambda}(d),
where the two holes occupy $xz$ and $yz$ orbitals and form a
total spin one. As can be inferred from the VCA grand potential shown in
Fig.~\ref{fig:omega_delta}, checkerboard AFM order arises as expected for
$\zeta = 0$. Orbital resolved hole densities in Fig.~\ref{fig:delta}(a) show that
the CF splitting $\Delta=1.5t$  is strong enough to completely fill the $xy$
orbital, so that we find pure spin-AFM without any excitonic character.  

Once SOC $\zeta$ sets it, the in-plane $x$ component of the spin rapidly becomes
favorable over the $z$ component, see the grand potentials in
Fig.~\ref{fig:omega_delta}. At the same time, hole density $n_{xy}^h$  in
$xy$ increases with SOC, see Fig.~\ref{fig:delta}(a), but
remains rather low with $n_{xy}^h\lesssim 0.2$ vs. 
$n_{xz/yz}^h \gtrsim 0.9$ for $\zeta \lesssim t$. However, for even stronger $\zeta
> 1.5 t$, magnetic symmetry breaking does not improve the grand potential,
see Fig.~\ref{fig:omega_delta}. This complete suppression of AFM order clearly
indicates a substantial role of SOC,  even though orbital polarization
continues to be substantial with  $n_{xy}^h\approx 0.4$ vs. 
$n_{xz/yz}^h \approx 0.8$. 

In the intermediate regime between the isotropic spin-AFM at $\zeta=0$ and the
nonmagnetic state at $\zeta > 1.5 t$, various linear combinations of $S$ and $L$ can have quite similar grand
potentials  as long as the ordered moment lies in the $x$-$y$ plane, see
Fig.~\ref{fig:omega_delta}. While $J_x$ dominates over a wide range of
$\zeta$, the small differences in $\Omega$ suggest that the exact form of the
order parameters is again far less important than its in-plane character.

More information about the intermediate magnetic state can be gained from the observables
plotted in Fig.~\ref{fig:delta}. While weight~(\ref{eq:wght_j}) is moved from $J=1$ and $J=2$ states
towards $J=0$ with growing $\zeta$, Fig.~\ref{fig:delta}(b) shows that
the $J=2$ states remain populated throughout, even in the PM regime at strong $\zeta
> 1.5\;t$. Nevertheless,  total angular momentum shows signatures of
excitonic magnetism for $\zeta \gtrsim 0.6t$:  when comparing the
gray symbols obtained in the PM state to the black ones of the AFM
state, one notices a weight transfer from $J=0$ to $J=1$, while weight in
$J=2$ is hardly affected. The $J=2$ states thus carry finite
weight, and contribute to orbital polarization, but are not involved in
the magnetic ordering. 

A clearer picture emerges when weight in eigenstates of $J$ is replaced
by weight in eigenstates of the effective onsite
Hamiltonian (\ref{eq:ham_onsite}), whose level structure is shown in Fig.~\ref{fig:delta_lambda}(a). 
Figure~\ref{fig:delta}(c) shows
that in the whole parameter regime, only the 
three lowest states $S_0$ and $D_{1/2}$ are relevant and fully capture a 
continuous transition from a spin-one via an excitonic magnet to the
PM phase. For 
moderate and strong $\zeta \gtrsim 0.7t$, magnetic ordering can be
described  in perfect analogy to  Fig.~\ref{fig:delta_0}(a), i.e.,
using the excitonic picture of weight transfer from the singlet
groundstate $S_0$ to the doublet $D_{1/2}$.


We now focus on the specific material example of Ca$_2$RuO$_4$ by
applying  the VCA to realistic
hopping parameters obtained from density-functional theory and projection onto Wannier
states. We use here the parameters derived in 
Ref.~\onlinecite{PhysRevLett.123.137204}, see its supplemental material for
details, but neglect very small hoppings. This yields $t_{xy}^{\textrm{NN}} = 0.2\;\textrm{eV}$,
$t_{xz}^{\textrm{NN}} = t_{yz}^{\textrm{NN}} = 0.137\;\textrm{eV}$,
$t_{xy}^{\textrm{NNN}}=0.1\;\textrm{eV}$ and  $t^o\approx
0.09\;\textrm{eV}$, as well as CF
$\Delta \approx 0.25\;\textrm{eV}$. Recent X-ray scattering experiments
have concluded that $U\approx 2\;\textrm{eV}$ and $J_H\approx
0.34\;\textrm{eV}$ and that oxygen covalency reduces  SOC from its free-ion value $\zeta \approx
0.16\;\textrm{eV}$ to $\zeta \approx0.13\;\textrm{eV}$~\cite{RIXS_Ca2RuO4_Gretarson19}.
As mentioned in the supplement of Ref.~\cite{PhysRevLett.123.137204},
double-counting corrections are here not needed. 

As seen in Fig.~\ref{fig:delta}(a), orbital occupations, with $n^h_{xy}\approx
0.2$, are very 
similar to those found matching values of $\zeta/t_{xz}$ \footnote{$t_{xz}^{\textrm{NN}} = t_{yz}^{\textrm{NN}}$ are
the two presumably most relevant hopping channels as they connect
approximately half-filled
orbitals}. Figure~\ref{fig:delta}(c) also shows that nearly all
weight is captured by the  onsite singlet $S_0$ and doublet $D_{1/2}$ and
Fig.~\ref{fig:delta}(b) reveals 
that magnetic ordering transfers weight from $J=0$ to $J=1$,
supporting an excitonic-magnetism description.  The main impact of the more
detailed kinetic energy is that we find the magnetic moment to prefer the
$x$-$y$ direction over the $x$ direction, in agreement with experiment.

\begin{figure}
  \includegraphics[width=0.48\columnwidth,clip]{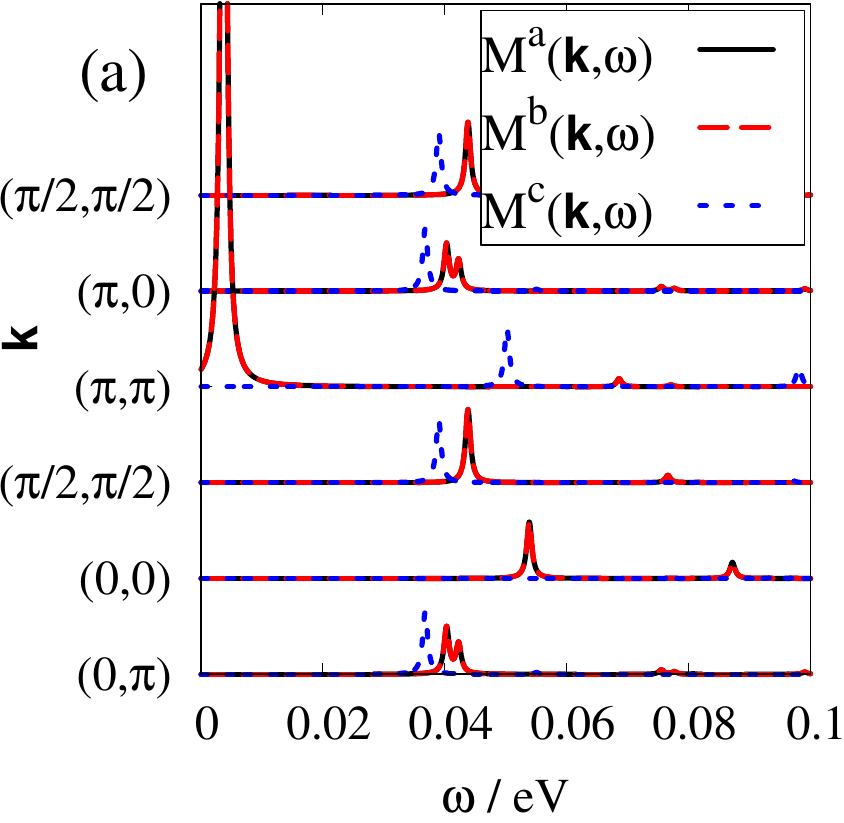}
  \includegraphics[width=0.48\columnwidth,clip]{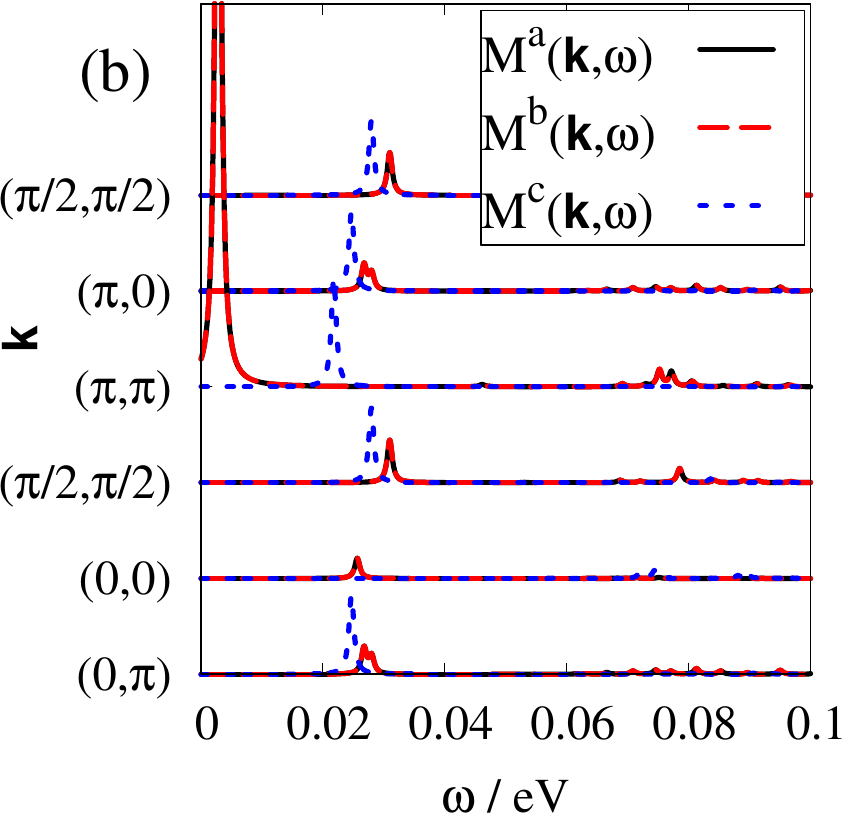}\\
  \includegraphics[width=0.48\columnwidth,clip]{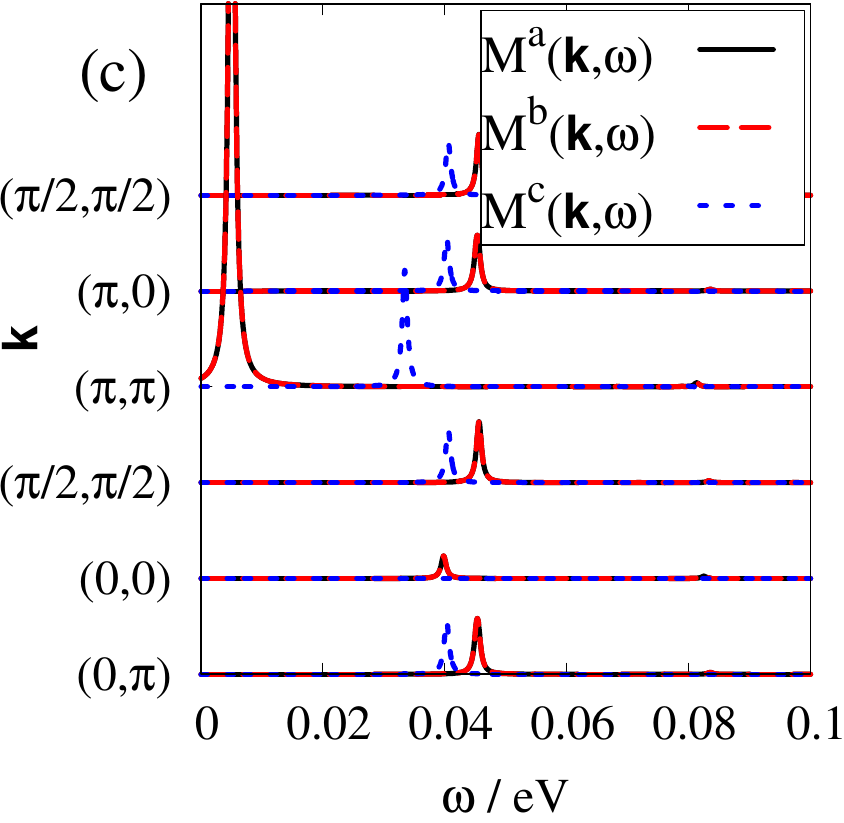}
  \includegraphics[width=0.48\columnwidth,clip]{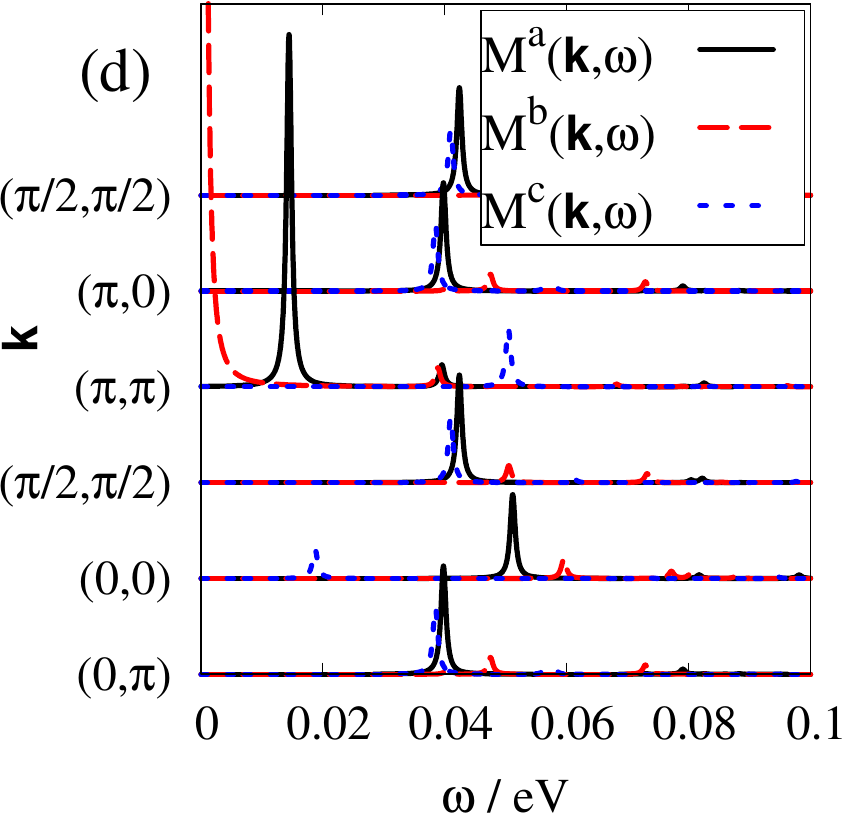}\\
\caption{Dynamic  magnetic structure factor (\ref{eq:Mkw}) for the
  Kugel-Khomskii model Eqs.~(\ref{eq:ham_onsite}) and (\ref{eq:HKKij})
  on an 8-site
  cluster (a) for parameters obtained to model Ca$_2$RuO$_4$ (see text), in
  particular $\zeta = 0.13\;\textrm{eV}$ and $\Delta = 0.25\;\textrm{eV}$ (b)
  the same but for reduced $\zeta =0.06\;\textrm{eV}$, (c) as in (a), but
  increased $\Delta = 0.5\;\textrm{eV}$, and (d) as in (a), but with additional magnetic 
  anisotropy $\delta \tfrac{1}{2}(S^{x}-S^y)^2$ favoring moments along the $b$ axis, with
  $\delta=-0.01\;\textrm{eV}$. Components $M^{a/b/c}(\vec{k},\omega)$  correspond
  to in-plane transverse ($M^a$), amplitude ($M^b$) and out-of-plane transverse ($M^c$) excitations. 
\label{fig:Skw}}
\end{figure}

\subsection{Magnetic excitations in Ca$_2$RuO$_4$}\label{sec:Skw}

The original and strongest hint for excitonic magnetism in Ca$_2$RuO$_4$
is the characteristic neutron-scattering spectrum with its clear maximum at
$(0,0)$ indicating pronounced $XY$ symmetry. While magnetic spectra
have been obtained using the VCA~\cite{Brehm_2010}, the method is not really
suitable for two-particle excitations. We accordingly use the observation that
the sum of all the weights $\langle J=0,1,2\rangle$ comes extremely close to one
for all parameter sets in Figs.~\ref{fig:delta_0}(a)
and~\ref{fig:delta}(b). This implies that the Hilbert space spanned by these
nine states with onsite $S=1$ and $L=1$ is sufficient to describe the ground state. 

We thus employ ED to calculate spectra for the the Kugel-Khomskii--like model obtained in second-order
perturbation theory in Sec.~\ref{sec:kk} for this limit. We use here the model parameters
derived for Ca$_2$RuO$_4$, but leave off inter-orbital hopping $t_o$ breaking
tetragonal symmetry. The spectrum in  Fig.~\ref{fig:Skw}(a) agrees well with neutron-scattering data for
Ca$_2$RuO$_4$~\cite{Higgs_Ru,PhysRevLett.115.247201} and reproduces
the salient feature of a maximum at $(0,0)$ and $\omega \approx
50\;\textrm{meV}$ as well as the dispersion at the zone boundary.

In order to probe the connection of the maximum at $(0,0)$ to SOC, we reduce
the latter from $\zeta =0.13\;\textrm{eV}$ to $\zeta =0.06\;\textrm{eV}$, which reduces hole density $n^h_{xy}$ from $n^h_{xy}
\approx 0.25$ to  $n^h_{xy} \approx 0.11$. At the same time, the maximum at
$(0,0)$ disappears, see  in Fig.~\ref{fig:Skw}(b), where one also
notes generally lower excitation energies. Similarly,
keeping $\zeta =0.13\;\textrm{eV}$ but increasing CF from $\Delta=0.25\;\textrm{eV}$ to
$\Delta=0.5\;\textrm{eV}$ enhances orbital polarization to $n^h_{xy} \approx
0.06$ and at the same time transforms the maximum at $(0,0)$ into a local minimum, see Fig.~\ref{fig:Skw}(c).

Finally, we introduce magnetic anisotropy $\delta (S^b)^2=\delta \tfrac{1}{2}(S^{x}-S^y)^2$,
with $S^{a/b} = \tfrac{1}{\sqrt{2}}(S^x\pm S^y)$ and $\delta = -10\;\textrm{meV}$, to
mimic the impact of $t^o$, which was found in VCA to prefer moments along the
$x-y$ direction. This reproduces the experimentally observed gap
of $\omega \approx 15\;\textrm{meV}$ at $(\pi,\pi)$~\footnote{Note that a finite
cluster cannot show spontaneous symmetry breaking.}. The
resulting spectrum is given in  Fig.~\ref{fig:Skw}(d) and while excitation
energies are slightly too large by $\approx 10\;\%$, it nevertheless
agrees well with the data reported in \onlinecite{Higgs_Ru}: In 
addition to the out-of-plane transverse mode $M^c$, the broken
in-plane symmetry splits the amplitude mode $M^b$ off from the
in-plane transverse mode $M^a$.

\subsection{Phase diagram}

\begin{figure}
  \includegraphics[width=\columnwidth]{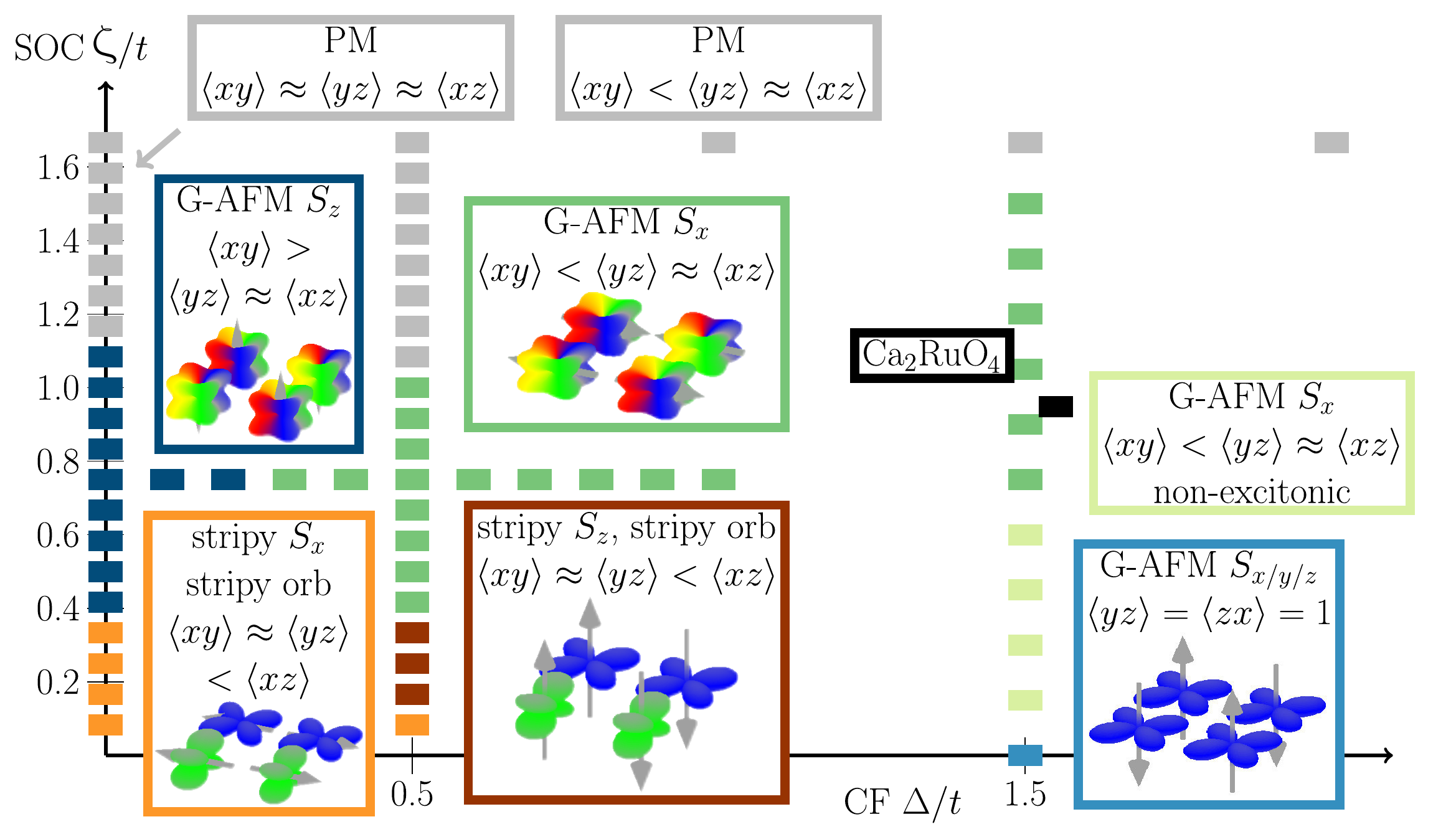}
  \caption{Approximate phase diagram for the simplified model with $t = t_{xy}^{\textrm{NN}} =
t_{xz}^{\textrm{NN}} =t_{yz}^{\textrm{NN}}$, $t_{xy}^{\textrm{NNN}} = t^{o} =
0$, $U=12.5t$ and $J_H=2.5t$. Expectation values $\langle\beta\rangle$ denote
orbital hole densities, with $\beta=yz,xz,xy$. Red and orange denote the stripy spin and
orbital order if Fig.~\ref{fig:delta_lambda}(b), with red designating a
preference for in-plane magnetic ordering and orange for out-of-plane
ordering. Dark blue designates the excitonic AFM state with out-of-plane magnetic
order and approximately equal hole densities in all orbitals. (For $\Delta=0$,
the layered geometry moves slightly more holes into the $xy$ orbital.)
Green refers to checkerboard $G$-AFM with in-plane ordered moment and
with orbital polarization pushing holes out of the $xy$ orbital. Light green indicates the regime where a description in
terms of a spin-one system is more appropriate, and darker green the excitonic
regime. The light blue data point indicates the perfectly isotropic
Heisenberg spin-one AFM without SOC and with a completely filled $xy$
orbital; gray indicates the PM phase. The approximate place of Ca$_2$RuO$_4$ in the phase diagram is
indicated. \label{fig:phases}}
\end{figure}

Going back to the simplified but more generic model without
nest-nearest--neighbor or interorbital hopping, we give a phase diagram in
Fig.~\ref{fig:phases}. For vanishing and small CF and SOC, we find the stripy
phase, with a small CF together with SOC favoring in-plane magnetism rather
than out-of-plane as found for $\Delta=0$. For stronger $\Delta$, magnetic
ordering is always checkerboard, but we use shades of green to indicate, where
the system can be more appropriately described as a spin-AFM, and where the
excitonic description is applicable. Note that there is a regime of SOC $\zeta
\approx t$ - $1.5t$, where applying a CF $\Delta$ can drive the system through
a quantum critical point between the PM and excitonic AFM phases. 

\section{Summary and Conclusions}\label{sec:conclusions}

We have used the VCA to investigate excitonic
magnetism stabilized by the interplay of superexchange and SOC in strongly correlated $t_{2g}$ orbitals. Without CF
splitting, the main role of SOC is to reduce orbital degeneracy, which in turn
changes the magnetic ordering vector from $(0,\pi)$ to $(\pi,\pi)$.  A similar transition from a state with complex
orbital order to an excitonic antiferromagnet has been found in one
dimension~\cite{PhysRevB.96.155111}. When
orbital degeneracy is from the outset reduced by a CF, SOC drives a transition
from an orbitally polarized $S=1$ AFM state, where the excitonic picture is not
applicable, via excitonic order coexisting with strong orbital polarization,
to a paramagnet where SOC suppresses superexchange and thus any magnetic ordering.

As an example, we focus on Ca$_2$RuO$_4$, which we model using hoppings and CF
strength derived from density-functional theory~\cite{PhysRevLett.123.137204} 
as well as interaction and SOC parameters inferred from
experiment~\cite{RIXS_Ca2RuO4_Gretarson19}. Magnetic excitations obtained
using ED for the corresponding  superexchange
model are found to closely match neutron-scattering data, in particular showing  the
tell-tale maximum at momentum $(0,0)$, and thus validate our microscopic
model. We have also shown that this maximum is clearly tied to the small hole
density $n^h_{xy} \approx 0.2 - 0.25$ remaining in the $xy$ model. When this
is further reduced by smaller SOC or larger CF, the maximum at $(0,0)$ becomes
a local minimum. 

Strong orbital polarization might suggest a description as an orbitally polarized $S=1$
system, in line with the interpretation of angle-resolved photon-emission--spectroscopy
data~\cite{hund_Ca2RuO4,PhysRevB.101.035115}. Indeed, the hole densities we
find would also support such a picture, as they clearly show that the holes
are overwhelmingly found in the $xz$ and $yz$ orbitals with $n^h_{xz/yz}
\approx 0.9$ vs. $n^h_{xy} \approx 0.2$. The excitonic character only becomes
apparent when looking at two-hole quantities like the total onsite angular
momentum or overlap of the onsite density matrix with the states depicted in
Fig.~\ref{fig:delta_lambda}(a), i.e. lowest eigenstates
$S_0$ and $D_{1/2}$ of the local Hamiltonian (\ref{eq:ham_onsite}). 
While magnetic ordering does not noticeably affect orbital densities,
it transfers weight from a singlet onsite ground state $S_0$ to doublet $D_{1/2}$, as expected for an
excitonic magnet. 
SOC and the remaining small
$xy$-hole density thus play a decisive role in the AFM state of 
Ca$_2$RuO$_4$, and we conclude Ca$_2$RuO$_4$ to realize excitonic magnetism despite its
pronounced orbital polarization. 

\begin{acknowledgments}
The authors acknowledge support by the state of Baden-W\"urttemberg through
bwHPC and via the Center for Integrated Quantum Science and Technology
(IQST). M.D. thanks KITP at UCSB for kind hospitality, this research was thus
supported in part by the National Science Foundation under Grant No. NSF
PHY-1748958. 
\end{acknowledgments}



%

\end{document}